\documentstyle[multicol,epsf,aps]{revtex}
\begin{document}
\title{Phase Transition in a Traffic Model with Passing}
\author{I. Ispolatov$^1$ and P. L. Krapivsky$^2$} 
\address{$^1$Department of Chemistry, Baker Laboratory, Cornell University,
  Ithaca, NY 14853}
\address{$^2$Center for Polymer Studies and   
  Department of Physics, Boston University, Boston, MA 02215}
\maketitle

\begin{abstract} 
\noindent  
We investigate a traffic model in which cars either move freely with
quenched intrinsic velocities or belong to clusters formed behind slower
cars.  In each cluster, the next-to-leading car is allowed to pass and
resume free motion.  The model undergoes a phase transition from a
disordered phase for the high passing rate to a jammed phase for the low
rate.  In the disordered phase, the cluster size distribution decays
exponentially in the large size limit.  In the jammed phase, the
distribution of finite clusters is independent on the passing rate, but
it accounts only for a fraction of all cars; the ``excessive'' cars form
an infinite cluster moving with the smallest velocity.  Mean-field
equations, describing the model in the framework of Maxwell
approximation, correctly predict the existence of phase transition and
adequately describe the disordered phase; properties of the jammed phase
are studied numerically.

\end{abstract}

\begin{multicols}{2} 
  
\section {Introduction}

Traffic flows on single-lane roads with no passing exhibit clustering
since queues of fast cars accumulate behind slow cars.  These clusters
form and grow even when car density is small.  The initial analysis of
cluster formation was carried out in the earlier days of traffic
theory\cite{newell}, and this subject continued growing ever
then\cite{eps,nag,benjam,krug,evans,wolf,kn,k}.  If passing is
introduced, the clusters may stop growing after reaching a certain size.
Indeed, previous work\cite{ep1,ep2,ep3} indicated that after a transient
regime a steady state is reached. The models of Refs.\cite{ep1,ep2,ep3}
assume that {\em any} car in a cluster can pass the leading car and the
passing rate is independent on the location of the car within the
cluster.  This is certainly an oversimplification of the everyday
traffic scenarios.  The complementary case when only the next-to-leading
cars can pass is also an idealization, yet it is closer to reality.
Below we show that the latter model is also richer phenomenologically as
it undergoes a dynamical phase transition.

We first comment on possible theoretical approaches. A mean-field theory
is the primary candidate, and we believe that it may be very good,
perhaps even exact, since clustering and passing mix positions and
velocities of the cars. The Boltzmann equation approach is an
appropriate mean-field scheme, and in our earlier work\cite{ep1,ep2} we
indeed used it. However, the present model, where only the
next-to-leading car is allowed to pass, is significantly more difficult
than the model\cite{ep1,ep2} where passing was possible for all cars.
Indeed, it appears impossible even to write down closed Boltzmann
equations for the distribution functions like $P(v,t)$ and $P_m(v,t)$,
the density of all clusters moving with velocity $v$, and the density of
clusters of $m$ cars, respectively.  Therefore our theoretical analysis
is performed in the framework of the Maxwell approach.  This scheme
simplifies ``collision'' terms by replacing the actual collision rates,
which are proportional to velocity difference of collision partners, by
constants. Despite this essentially uncontrolled approximation, the
Maxwell approximation is very popular in kinetic theory\cite{rl} and it
has already been used in traffic\cite{ep3}.

The important feature of our model is {\em quenched disorder}, which
manifests itself in the random assignment of intrinsic velocities. Road
conditions (construction zones, turns, hills, etc.)  present another
source of quenched randomness in real driving situations \cite{other},
which is ignored in our model.  Quenched disorder significantly affects
characteristics of many-particle systems, especially in low spatial
dimensions\cite{bg}. This general conclusion applies to the present
one-dimensional traffic model as we shall show below.

\section {Maxwell approximation}

We now formally define the model. Free cars move with {\em quenched}
intrinsic velocities randomly assigned from some distribution $P_0(v)$.
When a car or a cluster encounters a slower one, it assumes its velocity
and a larger cluster is formed. In every cluster, the next-to-leading
car is allowed to pass and resume driving with its intrinsic velocity.
The rate of passing is assumed to be a constant.  Thus clusters move and
aggregate deterministically, while passing is stochastic.  The system is
initialized by randomly placing single cars and assigning them
uncorrelated intrinsic velocities.
 
Within the Maxwell approach, the joint size-velocity distribution
function (the density of clusters of size $m$ moving with velocity $v$)
$P_m(v,t)$ obeys
\begin{eqnarray} 
\label{first} 
{\partial P_m(v,t)\over\partial t}
&=&\gamma (1-\delta_{m,1})[P_{m+1}(v,t)-P_m(v,t)]\nonumber\\ 
&+&\gamma \delta_{m,1}[N(v,t)+P_2(v,t)]-c(t)P_m(v,t)\nonumber \\
&+&\int_v^{\infty} dv' \sum_{i+j=m} P_i(v',t)P_j(v,t).
\end{eqnarray}
Here $\gamma $ is the passing rate, so terms proportional to $\gamma $
account for escape, while the rest describes clustering. The escape
terms are the same within Boltzmann and Maxwell approaches, and they are
actually {\em exact}.  The collision terms are mean-field by nature, and
they are different in the Boltzmann and Maxwell approaches.  For
instance, in the Boltzmann case, the integral term must involve $v'-v$.
Eqs.~(\ref{first}) also contain $c(t)$, the total cluster density
\begin{eqnarray} 
\label{ct} 
c(t)=\sum_{j\geq 1} \int_0^{\infty} dv\,P_j(v,t),
\end{eqnarray}
and $N(v,t)$, the density of clusters in which the next-to-leading car
has intrinsic velocity $v$.  This $N(v,t)$ causes the major trouble
since it cannot be expressed through $P_j(v,t)$.  One might try to close
Eqs.~(\ref{first}) by introducing $F_k(v,v',t)$, the density of clusters
moving with the velocity $v'$ whose $k^{\rm th}$ car has intrinsic
velocity $v$. Clearly, $N(v,t)=\int_0^v dv' F_2(v,v',t)$, and it appears
that equations for $F_k(v,v',t)$ are closed.  A more careful look,
however, reveals that the governing equation for $F_2(v,v',t)$ includes
three-velocity correlators.

Thus, at the first sight, the Boltzmann and Maxwell approaches appear to
be equally incapable of providing closed equations for the joint
size-velocity distribution function.  Still, the Maxwell framework has
an advantage that it does provide a closed description on the level of
the cluster size distribution.  Indeed, integrating Eqs.~(\ref{first})
over velocity and defining $P_m(t)\equiv \int_0^\infty dv\, P_m(v,t)$,
we find that the cluster size distribution $P_m(t)$ obeys
\begin{eqnarray} 
\label{simple} 
{d P_m\over dt}
=\gamma [P_{m+1}-P_m]-c\,P_m+{1\over 2}\sum_{i+j=m} P_iP_j
\end{eqnarray}
for $m\geq 2$, and
\begin{eqnarray} 
\label{simple1} 
{d P_1\over dt}
=\gamma [P_2-P_1+c]-c\,P_1.
\end{eqnarray}
Besides this formal derivation of Eqs.~(\ref{simple})--(\ref{simple1})
by direct integration of Eqs.~(\ref{first}), it is possible to obtain
these equations by enumerating all possible ways in which clusters
evolve. For instance, consider Eq.~(\ref{simple1}).  Collisions reduce
the density of single cars, and the collision rate is clearly equal to
$c(t)$, as it is velocity-independent in the framework of the Maxwell
approach.  The escape term in Eq.~(\ref{simple1}) is understood by
observing that the rate of return of single cars into the system is
equal to
\begin{eqnarray*} 
\gamma \left[2P_2+\sum_{j\geq 3} P_j\right]=
\gamma \left[P_2-P_1+c\right].
\end{eqnarray*}
Here $P_2(t)$ is singled out since passing transforms it into
two single cars while an escape from larger clusters produces only one
freely moving car.

Eqs.~(\ref{simple})--(\ref{simple1}) are closed.  Mathematically similar
equations were investigated previously in the context of the
aggregation-fragmentation model\cite{ps,barma}.  Therefore, we merely
present essential steps of the analysis.  Restricting ourselves to the
steady state and introducing notations $P_m=\gamma F_m$,
$c_\infty=\gamma F$, we recast Eqs.~(\ref{simple})--(\ref{simple1}) into
\begin{equation} 
\label{F} 
FF_m=F_{m+1}-F_m+\delta_{m,1}F+{1\over 2}\sum_{i+j=m} F_iF_j.
\end{equation}
These equations should be solved together with the constraints
$\sum_{m\geq 1}P_m=c_\infty$ and $\sum_{m\geq 1} mP_m=1$, i.e.,
\begin{equation} 
\label{sumF} 
\sum_{m\geq 1}F_m=F, \quad
\sum_{m\geq 1}mF_m=\gamma^{-1}.
\end{equation}
Note that the sum $\sum_{m\geq 1}mP_m(t)$ is obviously constant due to car
conservation. The constant is equal to the initial concentration $c_0$
as cars were initially unclustered. Here and below we always choose
$c_0=1$.

As in Ref.~\cite{ps}, we introduce the generating function
\begin{equation}
\label{genF}
{\cal F}(z)=\sum_{m\geq 1} (z^m-1) F_m.
\end{equation}
This generating function obeys
\begin{equation} 
\label{Fz} 
{1\over 2}\,{\cal F}^2+{1-z\over z}\,{\cal F}+{(1-z)^2\over z}\,F=0,
\end{equation}
with the solution
\begin{equation} 
\label{Fsol} 
{\cal F}(z)={z-1\over z}\left\{1-\sqrt{1-2zF}\right\}.
\end{equation}
The steady state solution (\ref{Fsol}) exists only when the generating
function is real for all the $0\leq z\leq 1$.  Hence, we require that
$2F\leq 1$.  Assuming that this condition is satisfied, we expand the
generating function in the powers of $z$ to obtain the steady state
concentrations:
\begin{equation}
\label{gammaF}
F_m={(2F)^m\over 2\sqrt{\pi}}\left\{{\Gamma\left(m-{1\over 2}\right)
\over \Gamma(m+1)}-2F\,
{\Gamma\left(m+{1\over 2}\right)\over \Gamma(m+2)}\right\}.
\end{equation}
This solution is still incomplete as we have not yet determined $F$.  To
find $F$ we use the sum rules (\ref{sumF}).  The first sum rule is
manifestly obeyed, while the second sum rule yields $\sum mF_m=d{\cal
  F}/dz|_{z=1} =1-\sqrt{1-2F}=\gamma^{-1}$.  Thus, $F={2 \gamma -1\over 2 \gamma ^2}$, 
which translates into $c_\infty=1-1/2\gamma$.

The steady state solution (\ref{gammaF}) exists for sufficiently high
passing rates, $\gamma\geq \gamma_c=1$.  For $\gamma>1$ and large $m$, 
the steady state
size distribution simplifies to
\begin{equation}
\label{Pmlarge}
P_m\simeq C m^{-3/2}\left[1-\left(1-\gamma^{-1}\right)^2\right]^m, 
\end{equation}
with $C=(4\pi)^{-1/2}\gamma^{-1}(\gamma - 1)^2$.  Apart from a power-law
prefactor, the size distribution exhibits an exponential decay, $P_m\sim
e^{-m/m^*}$, in the large size limit.  The characteristic size diverges,
$m^*\sim (\gamma-1)^{-2}$ as the passing rate approaches the critical
value $\gamma_c=1$.  In the critical case, the size distribution has a
power-law form
\begin{equation}
\label{Pmcrit}
F_m={3\over 4\sqrt{\pi}}\,{\Gamma\left(m-{1\over 2}\right)
\over \Gamma(m+2)} \sim m^{-5/2}.
\end{equation}

Let now the passing rate drops below the critical value ($\gamma<\gamma
_c$). Since $F$ cannot grow beyond $F_c=1/2$, it stays constant.
Therefore, $F_m$ is given by the same Eq.~(\ref{Pmcrit}) as in the
critical case, and the cluster size distribution reads $P_m=\gamma F_m$.
This implies $c_\infty=\gamma /2$, i.e., the sum rule $\sum
P_m=c_\infty$ is valid.  The second sum rule is formally violated: $\sum
mP_m=\gamma \ne 1$, i.e.  the cluster size distribution (\ref{Pmcrit})
accounts only for the fraction of all the cars present in the system.
The only possible explanation is the formation of an infinite cluster
that contains all the excessive cars.  The second sum rule then shows
that $1-\gamma $ of all the cars in the system are in this infinite
cluster.

Thus within the framework of the Maxwell approximation, our
traffic model displays a phase transition which separates the disordered
and jammed phases.  The steady state cluster concentration has different
dependence on the passing rate for these two phases:
\begin{equation}
\label{cinf}
c_\infty=\cases{1-1/2\gamma,      &$\gamma>1$;\cr
                \gamma/2,  &$\gamma<1$.}
\end{equation}
In the disordered phase, the size distribution decays exponentially in
the large size limit.  In the jammed phase, $P_m$ has a power law tail
and in addition there is an infinite cluster which contains the
following fraction of cars:
\begin{equation}
\label{I}
I=\cases{0,        &$\gamma>1$;\cr
         1-\gamma , &$\gamma<1$.}
\end{equation}
This phase transition is similar to phase transitions in driven
diffusive systems {\em without} passing\cite{benjam,krug,evans,wolf,k}
and to phase transitions in aggregation-fragmentation
models\cite{ps,barma,ziff}.  Also, the mechanism of the formation of the
infinite cluster has a strong formal analogy to Bose-Einstein
condensation \cite{evans,barma}.

Turning back to the joint size-velocity distribution (\ref{first}),
we note that the
lack of an {\em exact} expression for $N(v)$ in terms of $P_m(v)$ does not
mean the lack of a mean-field relation between these quantities.
Indeed, the density $N(v)$ of clusters in which the next-to-leading car
has intrinsic velocity $v$, can be written as
\begin{equation}
\label{mv}
N(v)=\int_0^v dv'\,\sum_{j\geq 2}P_j(v')\,\,
{C(v)\over \int_{v'}^\infty dv''\,C(v'')}.
\end{equation}
Here $\sum_{j\geq 2}P_j(v')$ is the density of ``true'' clusters (i.e.,
freely moving cars are excluded) moving with velocity $v'$.  Then,
$C(v)=P_0(v)-P(v)$ is the density of cars with intrinsic velocity $v$
which are currently slowed down, i.e., they are neither single cars, nor
cluster leaders.  Assuming that the velocities of cars inside clusters
are perfectly mixed, $C(v)/\int_{v'}^\infty dv''\,C(v'')$ gives the
probability density that the next-to-leading car in a true $v'$-cluster
has the velocity $v$.  The product form of Eq.~(\ref{mv}) reveals its
mean-field nature, which is consistent with the spirit of our
theoretical approach.  One can verify that Eq.~(\ref{mv}) agrees with
the sum rule $\int dv\,N(v)=\sum_{j\geq 2}P_j$, thus providing a useful
check of self-consistency.

Although Eqs.~(\ref{first}) with $N(v)$ given by (\ref{mv}) seem very
complex even in the steady-state regime, several conclusions can be
derived without getting their complete solution. We first
simplify Eqs.~(\ref{first}) by introducing  auxiliary functions
\begin{equation}
\label{Qmv}
Q_m(v)=\int_v^\infty dv'\,P_m(v').
\end{equation}
By inserting $P_m=-{dQ_m\over dv}$ into the Eqs.~(\ref{first}),
integrating resulting equations over $v$, and using the boundary
conditions $Q_m(v=\infty)=0$, we find
\begin{eqnarray} 
\label{Qm}
\gamma\left[Q_{m+1}(v)-Q_m(v)\right]&-&c\,Q_m(v)+{1\over 2}\sum_{i+j=m}
Q_i(v)Q_j(v)\nonumber\\
&=&\delta_{m1}\,\gamma\,q(v),
\end{eqnarray}
with
\begin{eqnarray} 
\label{qv}
q(v)=-Q_1(v)-\int_v^\infty dv'\,N(v').
\end{eqnarray}
Eqs.~(\ref{Qm}) are almost identical to the
Eqs.~(\ref{simple})--(\ref{simple1}), the velocity is just a parameter.
Consequently, we anticipate qualitatively similar results, 
$Q_m(v)\sim m^{-3/2}e^{-m/m^*}$, and
\begin{equation}
\label{Qmvas}
P_m(v)\sim m^{-1/2}e^{-m/m^*}.
\end{equation}
with the characteristic size $m^*(v,\gamma)$ dependent on both velocity and
passing rate. Our more rigorous generating function analysis, performed 
along the lines described above, confirms the asymptotic form (\ref{Qmvas}).

\section{Simulations}

Now let us examine what conclusions obtained within the Maxwell approach
are relevant for the original model.  We first re-derive the condition
for the phase transition in the complete velocity-dependent form. Let us
consider a system of reference with the origin moving with the slowest
car. We assume that the system is sufficiently large for the slowest car
to have negligible velocity.  We compare the total flux of cars
clustering behind this slowest car, $\sum mP_m\langle v\rangle_m$, to
the rate of escape, $\gamma $. Here $\langle v\rangle_m$ is an average
velocity of a cluster of size $m$. When the rate of escape becomes less
than the rate of accumulation of the cars, the cluster behind the
slowest car (analog of the ``infinite cluster'' for finite systems)
grows to remove the excessive cars from the system.  Hence, the phase
transition point $\tilde \gamma_c$ is defined as
\begin{equation}
\label{phtr}
\sum_{m\geq 1} mP_m\langle v\rangle_m = \tilde \gamma_c.
\end{equation}
For the Maxwell model, where $\langle v\rangle_m=1$ for all $m$, 
Eq.~(\ref{phtr}) reduces to $\sum mP_m=\gamma_c=1$ as obtained above.
Since large clusters usually form behind slow cars, $\langle
v\rangle_m$ is a decreasing function of the cluster size $m$. In
particular, $\langle v\rangle_m$ is always smaller than the average car
velocity $\langle v\rangle$, implying $\tilde \gamma_c<1$.  

For a rough estimate of $\langle v\rangle_m$, consider a cluster of $m$
cars and {\em assume} that intrinsic velocities of the cars in the
cluster are independent. The leading car has the minimal velocity, so
the size-velocity distribution reads
\begin{equation}
\label{extrem}
P_m(v)\approx mP_0(v)\left[\int_v^\infty dv'\,P_0(v')\right]^{m-1}P_m.
\end{equation}
For concreteness, let us consider intrinsic velocity distributions which
behave algebraically near the lower cutoff, $P_0(v)\sim v^\mu$ as $v\to
0$. Then for large clusters we get
\begin{equation}
\label{extr}
P_m(v)\sim P_m\,\exp\left(-mv^{\mu+1}\right).
\end{equation}
This implies that the average cluster velocity $\langle v\rangle_m$
scales with $m$ according to $\langle v\rangle_m \sim m^{-1/(\mu+1)}$,
and hence $\tilde \gamma_c \sim \sum m^{\mu/(\mu+1)}P_m$.  We conclude
that the phase transition does exist in the original model, although its
location is shifted towards lower passing rate compared to the Maxwell
model prediction. This shift is especially significant for small $\mu$
($\mu>-1$ from the normalization requirement).

To check the relevance of other predictions of the Maxwell approach, we
performed molecular dynamics simulations.  We place $N=20000$ single
cars onto the ring of length $L=N$, so that the average car density is
equal to one. Initial positions and velocities of cars were assigned
randomly.  We considered linear $P_0(v)={8\over 9}\,v$ ($0<v<3/2$),
exponential $P_0(v)=e^{-v}$, and $P_0(v)=(2\pi v)^{-1/2}e^{-v/2}$
velocity distributions, which correspond to $\mu=1, 0, -1/2$ for the
small-$v$ asymptotics.  All these three distributions have the average
velocity equal to one.

In Fig.~1, we plot $\ln [m^{3/2}P_m]$ vs. $m$ for the above three
velocity distributions. We take $\gamma=1$ which, as we concluded
before, lies above the phase transition point $\tilde\gamma_c$.  We
expect the system to be in the disordered phase with $P_m$ being
expressed by Eq.~(\ref{Pmlarge}).  For the exponential and $P_0(v)=(2\pi
v)^{-1/2} e^{-v/2}$ intrinsic velocity distributions, there is a good
agreement with the prediction of the Maxwell model (\ref{Pmlarge}); for
the linear velocity distribution, there are some deviations for small
$m$, but for large $m$ the agreement is satisfactory. The slopes of the
plots decrease with $\mu$.  Taking into account that at the point of the
phase transition the slope equals to zero, this qualitatively confirms
that $\tilde\gamma_c$ gets smaller when $\mu$ decreases.

\begin{figure}
  \centerline{\epsfxsize=8cm \epsfbox{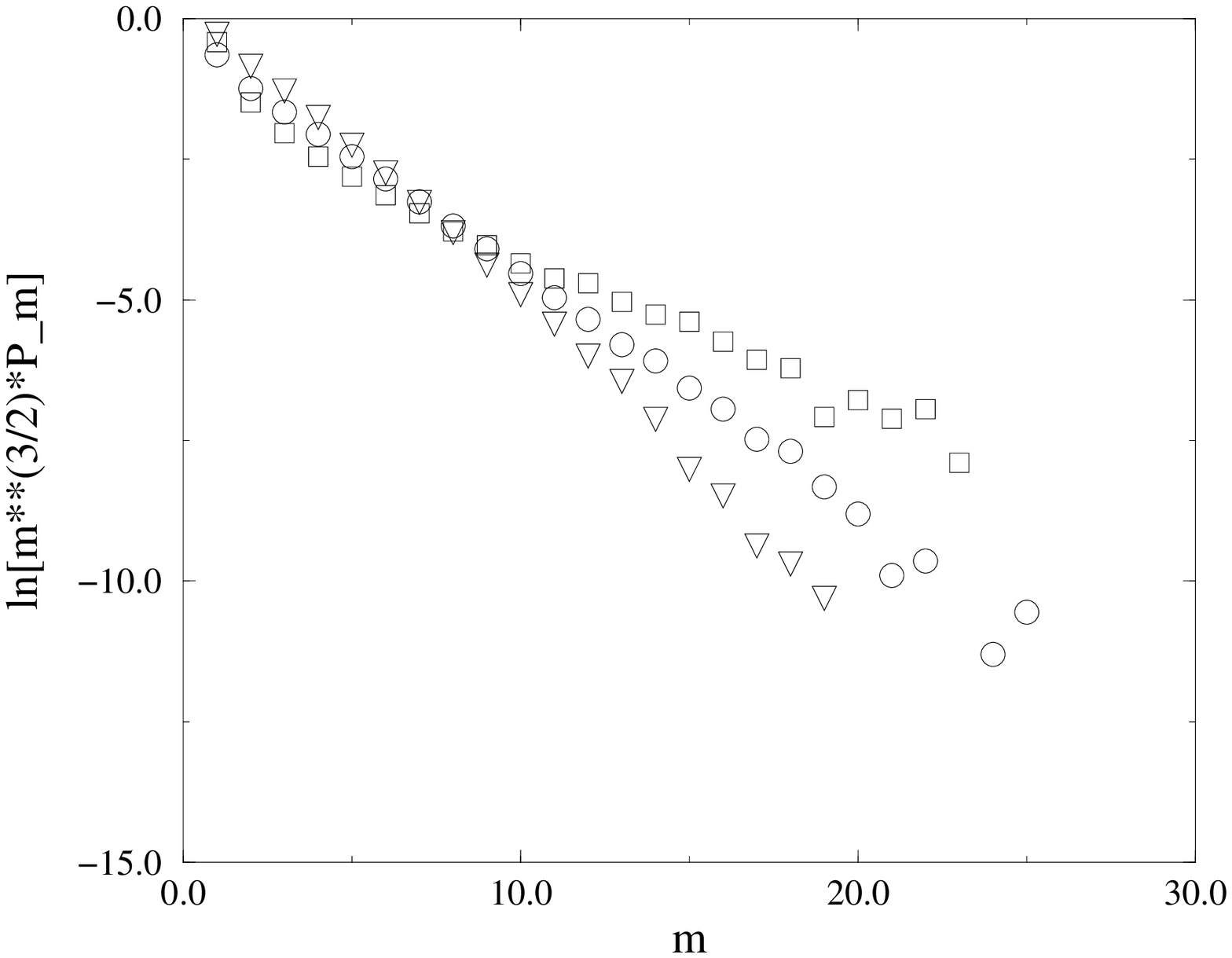}} {{\small {\bf Fig.~1}.
      Plot of $\ln\left[m^{3/2}P_m\right]$ vs.  cluster size $m$ in the
      high passing rate regime ($\gamma=1$) for linear ($\Box $), exponential
      (o), and $P_0(v)=(2\pi v)^{-1/2}e^{-v/2}$ ($\nabla$) initial
      velocity distributions.}}
\end{figure}

Plots of $P_m$ vs. $m$ for intrinsic velocity distributions
$P_0(v)=e^{-v}$ and $P_0(v)=(2\pi v)^{-1/2}e^{-v/2}$, with passing rate
$\gamma =0.005$ well below the phase transition point, are shown in
Fig.~2a and Fig.~2b, respectively.  The cluster size distribution
clearly consists of two regions: almost power-law tail for smaller $m$
and several separate peaks for larger $m$. These peaks correspond to the
fluctuating size of the infinite cluster, while the power-law tail
describes the regular part of $P_m$.  The apparent exponent $\tau$ of
the power-law region $P_m \sim m^{-\tau}$ slightly varies for different
passing rates and $P_0(v)$, though it remains confined between 3/2 and 2. It
is definitely different form the value 5/2, predicted by the Maxwell
model (\ref{Pmcrit}).  The measured values of $\tau$ would make the
total amount of cars in the system divergent, $\sum mP_m \rightarrow
\infty$, so the power-law region ends with an exponential cutoff at
large $m$. 

We now comment on the relationship of our model to earlier work. On the
mean-field level, our model is similar to the models of
Refs.\cite{ps,barma}.  On the level of the process, our model reminds an
asymmetric conserved-mass aggregation model (ASCMAM)\cite{barma} where
clusters undergo asymmetric diffusion, aggregation upon contact, and
chipping (single-particle dissociation).  Of course, our model is
continuum while the ASCMAM is the lattice model. More substantial
difference between the two models lies in the nature of randomness -- in
our model intrinsic velocities are quenched random variables, while in
the ASCMAM dynamics is the only source of randomness.  Nevertheless, the
phenomenology of the two models appears to be quite similar.  In
particular, the ASCMAM undergoes a phase transition, and in the jammed
phase, the cluster size distribution exibits a power law decay with the
exponent close to 2\cite{barma}.  We should stress that in the jammed
phase, we have not reached a scale-free critical state which must have
the exponent $\tau\geq 2$.  Maybe quenched randomness does not allow the
system to organize itself into a truly normalizable critical state.
Other possible explanation relies on large fluctuations in disordered
systems, i.e., our system was not large enough to ensure self-averaging.

\begin{figure}
  \centerline{\epsfxsize=8cm \epsfbox{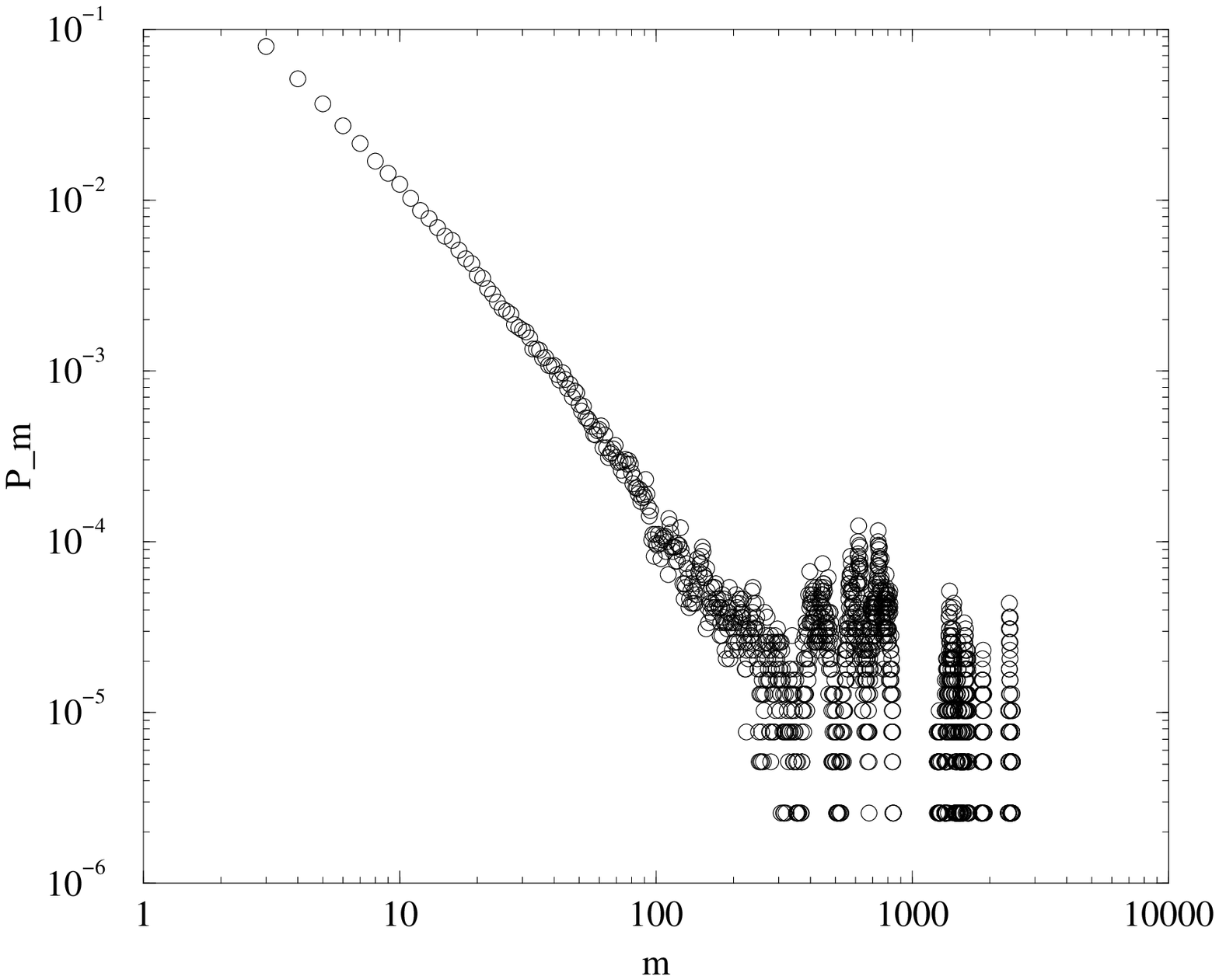}} {{\small {\bf
        Fig.~2a}.  Plot of the steady state cluster size distribution
      $P_m$ in the low passing rate regime ($\gamma =0.005$) for the
      exponential initial velocity distribution.}}
\end{figure}

\begin{figure}
  \centerline{\epsfxsize=8cm \epsfbox{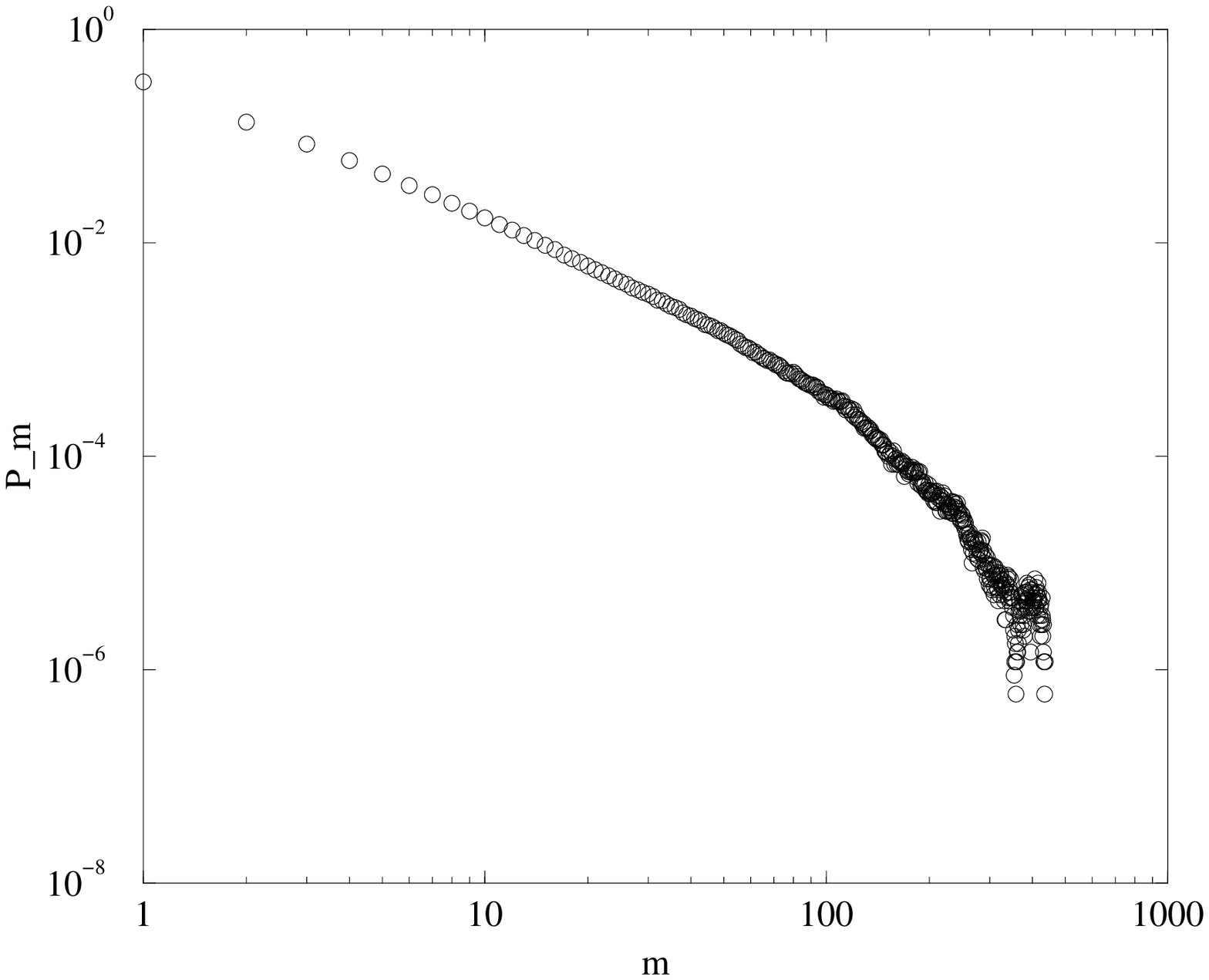}} {{\small {\bf
        Fig.~2b}.  Plot of the steady state cluster size distribution
      $P_m$ in the low passing rate regime ($\gamma =0.005$) for the
      $P_0(v)=(2\pi v)^{-1/2}e^{-v/2}$ initial velocity distribution.}}
\end{figure}

\section{Conclusion}

In this paper, we have investigated the model of traffic that involves
clustering and passing of the next-to-leading car. Despite the fact that
it is one of the simplest (if not the simplest) possible continuous
model of one-lane traffic with passing, the model has rich kinetic
behavior. Depending on the passing rate $\gamma$ the system organizes
itself either into disordered phase where density of large clusters is
exponentially suppressed, or into the jammed phase, where the cluster
size distribution becomes independent on $\gamma$ and the infinite
cluster is formed.  Within the framework of Maxwell approach, which
plays the role of the mean-field theory in the present context, we have
shown that the model admits an analytical solution.  We have argued that
the Maxwell approach correctly predicts the existence of the phase
transition and adequately describes the properties of the disordered
phase which arises when the passing rate is high.  For the jammed phase,
the Maxwell approach correctly predicts that the system stores excessive
cars in the infinite cluster and organizes itself into some kind of a
critical state.  However, the Maxwell approach cannot quantitatively
describe other properties of the jammed phase. It would be interesting
to design a more accurate theoretical approach which would allow to
probe the characteristics of the low passing rate regime analytically.
Some properties of the jammed state appear similar to the properties of
the jammed state of a lattice model of Ref.\cite{barma} which includes
an asymmetric lattice diffusion, aggregation, and fragmentation. It
would be interesting to gain a deeper understanding of the relationship
between these models, and whether the quenched disorder is the main
source of difference.

\medskip\noindent We are thankful to E.~Ben-Naim and S.~Redner for
discussions, and to NSF and ARO for support of this work.

\end{multicols}
\end{document}